\documentclass[twocolumn,showpacs,preprintnumbers,amsmath,amssymb,aps]{revtex4}

\usepackage{graphicx}
\usepackage{dcolumn}
\usepackage{bm}

\begin{document}
\preprint{APS/123-QED}
  \title{Aging and Rejuvenation with Fractional Derivatives}
\author{Gerardo Aquino$^{1}$, Mauro Bologna $^{1}$, Paolo Grigolini$^{1,2,3}$, Bruce
J. West $^{4}$}

\affiliation{$^{1}$Center for Nonlinear Science, University of North Texas,
P.O. Box 311427, Denton, Texas 76203-1427\\
$^{2}$Dipartimento di Fisica dell'Universit\`{a} di Pisa and INFM, via
Buonarroti 2, 56127 Pisa, Italy\\
$^{3}$Istituto dei Processi Chimico Fisici del CNR Area della Ricerca di
Pisa, Via G. Moruzzi 1, 56124 Pisa, Italy\\
$^{4}$ Mathematics Division, Army Research Office, Research Triangle Park,
NC 27709, USA }
\date{\today}

\begin{abstract}
We discuss a dynamic procedure that makes the fractional derivatives emerge
in the time asymptotic limit of non-Poisson processes. We find that
two-state fluctuations, with an inverse power-law distribution of waiting
times, finite first moment and divergent second moment, namely with the
power index $\mu $ in the interval $2<\mu <3$, yields a generalized master
equation equivalent to the sum of an ordinary Markov contribution and of a
fractional derivative term. We show that the order of the fractional
derivative depends on the age of the process under study. If the system is
infinitely old, the order of the fractional derivative, $ord$, is given by $%
ord=3-\mu $. A brand new system is characterized by the degree $ord=\mu -2$.
If the system is prepared at time $-t_{a}<0$ and the observation begins at
time $t=0$, we derive the following scenario. For times $0<t<<t_{a}$ the
system is satisfactorily described by the fractional derivative with $%
ord=3-\mu $. Upon time increase the system undergoes a rejuvenation process
that in the time limit $t>>t_{a}$ yields $ord=\mu -2$. The intermediate time
regime is probably incompatible with a picture based on fractional
derivatives, or, at least, with a mono-order fractional derivative.
\end{abstract}
\maketitle


\section{Introduction}

The fractional calculus has recently received a great deal of attention in
the physics literature, through the publications of books \cite{hilfer,brucemauro}, review articles \cite{metzlerklafter,zaslavsky}, as well
as an ever increasing number of research papers, some of which are quoted
here \cite{researchpaper1,lutz,gorenflo,bischero,researchpaper2,researchpaper3,researchpaper4,sokolov}%
. The blossoming interest in the fractional calculus is due, in part, to the
fact it provides one of the dynamical foundations for fractal stochastic
processes \cite{brucemauro,zaslavsky}. The adoption of the fractional
calculus by the physics community was inhibited historically because there
was no clear experimental evidence for its need. The disciplines of
statistical physics and thermodynamics were thought to be sufficient for
describing complex physical phenomena solely with the use and modifications
of analytic functions. This view was supported by the successes of such
physicists as Lars Onsager, who through the use of simple physical arguments
was able to relate apparently independent transport processes to one
another, even though these processes are associated with quite different
physical phenomena \cite{onsager}. His general arguments rested on three
assumptions: 1) microscopic dynamics have time-reversal symmetry; 2)
fluctuations of the heat bath decay at the same rate as do macroscopic
deviations from equilibrium and 3) physical systems are aged. We refer to
assumption 2 as the Onsager Principle and show that it is tied up with
assumption 3.

Onsager's arguments focused on a system that is in contact with a heat bath
sufficiently long that the bath has come to thermal equilibrium and
consequently the system is aged. In statistical physics we know that the
bath is responsible for both fluctuations and dissipation, and if the
fluctuations are white the regression of perturbations of the bath to
equilibrium is instantaneous. This means that the energy absorbed from the
system of interest by the bath, through macroscopic dissipation, is
distributed over the bath degrees of freedom on a very much shorter time
scale than the relaxation time of the system. This property is summarized in
the well known fluctuation-dissipation theorem, which has even been
generalized to the case where the fluctuations in the bath do not regress
instantaneously \cite{kubo}.

The dynamics of the physical variables to which the Onsager Principle apply
are described by two different kinds of equations: 1) the Langevin equation,
a stochastic differential equation for the dynamical variable and 2) the
phase space equation for the probability density. Two distinct methods have
been developed to describe the phase space evolution of the probability
density: the master equation introduced by Pauli and the continuous time
random walk (CTRW) approach of Montroll and Weiss \cite{montrollweiss}. The
CTRW formalism describes a random walk in which the walker pauses after each
jump for a sojourn specified by a waiting time distribution function. It was
shown by Bedeaux {\em et al.} \cite{bedeaux} that the Markov master equation
is equivalent to CTRW if the waiting time distribution is Poissonian.
However, when the waiting time distribution is not exponential, the case we
consider here, the equivalence between the two approaches is only maintained
by generalizing to the non-Markov master equation, the so-called GME  \cite
{kenkre}. Recently, Metzler  \cite{metzler} argued that the GME unifies the
fractional calculus and CTRW.

Allegrini {\em et al.}  \cite{gerardo} have shown that to create a master
equation compatible with the Onsager Principle requires that the system be
entangled with the heat bath in such a way that the bath does not regress to
equilibrium infinitely fast. The system-bath entanglement is the result of a
rearrangement process that may take infinitely long to complete, leading to
the generalized master equation (GME) of Kenkre {\em et al.}  \cite{kenkre}.
Allegrini {\em et al.} \cite{gerardo} were concerned with how to make the GME
stationary and therefore compatible with the Onsager Principle. Herein, we
extend that discussion to include the connection with both the fractional
calculus and the nonstationary condition.

For simplicity we restrict our discussion to the case of a two-site system
coupled to a heat bath, a problem closely related to the quantum problem of
decoherence of a qubit, due to coupling to the environment. The connection
of the CTRW, a two-state non-Markov master equation and the fractional
calculus has been explored by Sokolov and Metzler  \cite{sokolov}. Herein we
go beyond their discussion and investigate the relationships among the
fractional calculus, a particular generalization of the Onsager Principle,
and one concept of aging. We also show that the result of Sokolov and
Metzler, which refers to the young state, is an attractor for all the
systems that are partially aged, and not infinitely aged. We call this
process rejuvenation.

\subsection{Beyond the Onsager Principle}

We approach the subject of fractional derivatives from a perspective similar
to that of Sokolov and Metzler  \cite{sokolov}. More specifically, we address
the problem of the connection between the GME  \cite{kenkre} and the
stationary version of the CTRW  \cite{montrollweiss}. The GME considered by
Allegrini {\em et al.}  \cite{gerardo} is the two-site version of the
following equation

\begin{equation}
\frac{d}{dt}{\bf p}(t)=-\int_{0}^{t}\Phi (t-t^{\prime }{){\bf {Kp}}}
(t^{\prime })dt^{\prime },  \label{kenkre}
\end{equation}
where ${\bf p}\left( t\right) $ is the $m$-dimensional population vector of $%
m$ sites, ${\bf K}$ is a transition matrix between the sites and $\Phi (t)$
is the memory kernel. The CTRW prescription for this process yields

\begin{equation}
{\bf {p}}\left( t\right) =\sum_{n=0}^{\infty }\int_{0}^{t}dt^{\prime }\psi
_{n}\left( t^{\prime }\right) \Psi \left( t-t^{\prime }\right) {\bf {M}}^{n}%
{\bf {p}}\left( 0\right) .  \label{ctrw1}
\end{equation}
Note that $\psi _{n}\left( t\right) $ is the probability that $n$ jumps
occurred and that the last jump took place at time $t=t^{\prime },$ implying
the renewal theory relation

\begin{equation}
\psi _{n}\left( t\right) =\int_{0}^{t}\psi _{n-1}\left( t-t^{\prime }\right)
\psi _{1}\left( t^{\prime }\right) dt^{\prime } ,  \label{ctrw2}
\end{equation}
where $\psi _{1}\left( t\right) $ is the waiting time distribution function $%
\psi \left( t\right) $ introduced into CTRW and $\psi _{0}\left( t\right)
=\delta \left( t\right) .$ While ${\bf M}$ is the transition matrix
connecting the sites after one jump has occurred, the probability that no
jump occurs in the time interval $(0,t)$ is

\begin{equation}
\Psi \left( t\right) =\int_{t}^{\infty }\psi \left( t^{\prime }\right)
dt^{\prime }.  \label{ctrw3}
\end{equation}
The waiting time distribution function and the memory kernel can be related
to one another by taking the Laplace transform of the GME (\ref{kenkre}) and
the CTRW (\ref{ctrw1}). This comparison, after some algebra  \cite{gerardo},
yields

\begin{equation}
\hat \Phi(u) {\bf K} = \frac{u \hat \psi(u)}{1 - \hat \psi(u)} ({\bf M} - {\bf I}),
 \label{ctrw4}
\end{equation}
where ${\bf I}$ is the $m\times m$ unit matrix and the Laplace transform of
the function $f\left( t\right) $ is $\hat{f}\left( u\right) $.
Here, as in Allegrini {\em et al.} \cite{gerardo}, we limit our discussion to the
two-state case where

\begin{equation}
{\bf M}=\left(
\begin{array}{cc}
0 & 1\\
1 & 0\\
\end{array}\right)
\end{equation}
and

\begin{equation}
\label{ctrw6}
{\bf K}=\left(
\begin{array}{cc}
1 & -1\\
-1 & 1\\
\end{array}\right),
\end{equation}thereby reducting (\ref{ctrw4}) to

\begin{equation}  \label{brandnew}
\hat{\Phi }\left( u\right) =\frac{u\hat{\psi }%
\left( u\right) }{1-\hat{\psi }\left( u\right) }.
\end{equation}
This relation between the Laplace transform of the memory kernel and the
Laplace transform of the waiting time distribution function was first
obtained by Kenkre {\em et al.}  \cite{kenkre} and is reviewed by Montroll
and West  \cite{montroll}.

In the case when the lattice has only two sites, a left and a right site,
the random walker corresponds to a dichotomous signal $\xi $, with the
values $\xi (t)=-1$, for the left site, and $\xi (t)=1$, for the right site.
For the sake of simplicity, we assume the two states to have the same
statistical weight. Also in the two-state CTRW, if we adopt a discrete time
representation, the motion of the random walker corresponds to a symbolic
sequence $\left\{ \xi \right\} $, with the form $\left\{
+++++++-+--++++-------....\right\} ,$ which shows a significant
persistence of both states. The waiting time distribution $\psi \left(
t\right) $ is the distribution of the patches filled with either +'s or -'s.
We assume symmetry between the two states and a finite first moment of $\psi
\left( t\right) $ making it possible for us to define the autocorrelation
function for the fluctuations $\xi \left( t\right) $

\begin{equation}
\Phi _{\xi }(t)=\frac{<\xi (0)\xi (t)>}{<\xi ^{2}>},
\label{correlationfunction}
\end{equation}
because the process is stationary in time \cite{renewal}.

In his original work Onsager considered the case of a macroscopic
fluctuation that regresses to equilibrium through the phenomenological
equations of motion. Here we adopt an extension of the Onsager Principle
made by Allegrini {\em et al.}  \cite{gerardo} to the case of two states,
using the probability of the random walker being in state $i=1,2$, $%
p_{i}\left( t\right) $, at time $t$,

\begin{equation}
\Phi _{\xi }(t)=\frac{p_{1}\left( t\right) -p_{2}\left( t\right) }{%
p_{1}\left( 0\right) -p_{2}\left( 0\right) }.  \label{ctrw8}
\end{equation}
The GME cannot provide information about the autocorrelation function of the
fluctuations, but only about the probability of occupying a given state $%
i=1,2$, $p_{i}\left( t\right) $, at time $t.$ Thus, given that the state is
initially out of equilibrium $p_{1}\left( 0\right) -p_{2}\left( 0\right)
\neq 0$, the generalized Onsager Principle given by (\ref{ctrw8}) is the
only way to relate the bath auto-correlation function to the dynamics of the
system. Assuming a regression to equilibrium in such a way as the retain (%
\ref{ctrw8}) we obtain from the GME (\ref{kenkre}) using the coupling matrix
(\ref{ctrw6}):

\begin{equation}
\frac{d\Phi _{\xi }\left( t\right) }{dt}=-2\int\limits_{0}^{t}\Phi \left(
t-t^{\prime }\right) \Phi _{\xi }\left( t^{\prime }\right) dt^{\prime }.
\label{58}
\end{equation}
Thus, the Laplace transform of the auto-correlation function can be related
to the Laplace transform of the memory kernel by

\begin{equation} \hat{\Phi }_{\xi }\left( u\right) =\frac{1}{u+2\hat{\Phi }\left( u\right) }.  \label{lapcorr}
\end{equation}

The authors of Ref.  \cite{gerardo} studied the problem of establishing a
complete correspondence between the GME and renewal theory, characterized by
the waiting time distribution 
\begin{equation}
\psi (t)=(\mu -1)\frac{T^{\mu -1}}{(T+t)^{\mu }},
\label{distributiondensity}
\end{equation}
with $\mu >2$ to fit the stationary condition. These authors determined that this problem could
be solved by expressing the  CTRW  in stationary form, resulting in the GME
memory kernel:

\begin{equation}  \label{gerardoisagenius}
\hat{\Phi }\left( u\right) =\frac{u\left( 1-\hat{\psi }\left( u\right) \right) }{-2\left( 1-\hat{\psi }\left(
u\right) \right) +u\left( 1+\hat{\psi }\left( u\right) \right)
\tau } , 
\end{equation}
where $\tau $ is the average waiting time

\begin{equation}
\tau =\int_{0}^{\infty }t\psi \left( t\right) dt=\frac{T}{\mu -2}.
\label{meanvalue}
\end{equation}
The form of the memory kernel given by (\ref{gerardoisagenius}) is consistent with the
equation of motion for the auto-correlation function (\ref{58}), and
consequently, Eq. (\ref{gerardoisagenius}) is equivalent to 
\begin{equation}
\hat{\Phi}(u)=\frac{1}{2}(\frac{1}{\hat{\Phi}_{\xi }(u)}-u).
\label{mauroisageniustoo}
\end{equation}

We have to remind the reader that the stationary auto-correlation function of $\xi$ is not related directly to $\psi(t)$. 
 Zumofen and Klafter  \cite{fundamental} provided a prescription for
deriving the corresponding  equilibrium auto-correlation function of $\xi$ from $\psi (t)$.
Their result rests on the observation that $\psi (t)$ is an experimental
function, evaluated by observing the time duration of the two states. The
connection with renewal theory is established by assuming that the time
duration of a state is determined by two processes, one is the extraction of
a random number from a theoretical inverse power law distribution, $\psi
^{*}(\tau )$, with the same power index $\mu $, and the other is a coin
tossing procedure that determines the sign of this laminar region. Thus, a
given experimental sojourn time in one of the two states may correspond to
an arbitrarily large number of drawings and coin tossings. Renewal theory is
used to relate the auto-correlation function $\Phi _{\xi }(t)$ to the
waiting time distribution function $\psi^{*} \left( t\right) $. 
In fact, from the renewal theory  \cite{renewal} we obtain the following
important result: 
\begin{equation}
\Phi _{\xi }(t)=\frac{1}{\tau ^{*}}\int_{t}^{\infty }(t^{\prime }-t)\psi
^{*}(t^{\prime })dt^{\prime },  \label{geiselcrucial1}
\end{equation}
where $\tau ^{*}$ is the mean waiting time of the $\psi ^{*}(t)$%
-distribution density. 
It is interesting to notice that this equation
implies that the second derivative of the auto-correlation function is
proportional to $\psi ^{*}(t)$, 
\begin{equation}
\frac{d^{2}}{dt^{2}}\Phi _{\xi }(t)=\frac{\psi ^{*}(t)}{\tau ^{*}}.
\label{geiselcrucial2}
\end{equation}

In Section 2 the departure point of our calculations is given by the auto-correlation function $\Phi_{\xi}(t)$ of Eq. (\ref{geiselcrucial1}). In this case is convenient to assign to this equilibrium auto-correlation  function a simple analytical form. This is done as follows. We assign the form of Eq. (\ref{distributiondensity}) to $\psi^{*}(t)$, thereby writing
\begin{equation}
\psi ^{*}(t)=(\mu -1)\frac{{T^{*}}^{\mu -1}}{(t+T^{*})^{\mu }}.  \label{star}
\end{equation}
This makes it possible for us to write $\tau^{*}$ as follows
\begin{equation}
\tau^{*} =\int_{0}^{\infty }t\psi^{*} \left( t\right) dt=\frac{T^{*}}{\mu -2}.
\label{meanvaluestar}
\end{equation}

With the choice of Eq. (\ref{star}) for $\psi^{*}(t)$,  the autocorrelation function $\Phi_{\xi}(t)$ of Eq. (\ref{geiselcrucial1}) gets
the attractive analytical form 
\begin{equation}
\Phi _{\xi }(t)=\frac{{T^{*}}^{\beta }}{(T^{*}+t)^{\beta }},
\label{autocorrelationfunction}
\end{equation}
where 
\begin{equation}
\beta \equiv \mu -2.  \label{beta}
\end{equation}
Thus, in the case $\mu <3$, the auto-correlation function of the fluctuations
is not integrable.

Zumofen and Klafter  \cite{fundamental}, in addition to explaining with clear
physical arguments the connection between $\psi (t)$ and $\psi ^{*}(t)$,
established that the Laplace transforms of the two functions are related one
to the other by 
\begin{equation}
\hat{\psi}^{*}(u)=\frac{2\hat{\psi}(u)}{1+\hat{\psi}(u)}.  \label{connection}
\end{equation}
This important relation allows us to establish a connection between $\tau$ and $\tau^{*}$, which turns out to be
\begin{equation}
\label{fromtautotaustar}
\tau = 2 \tau^{*}.
\end{equation}

The equivalence between Eq. (\ref{gerardoisagenius}) and Eq. (\ref{mauroisageniustoo}) rests on the key property of Eq. (\ref{connection}). We note that if we make the choice of Eq. (\ref{distributiondensity}) then the waiting time distribution  $\psi^{*}$ loses the simple analytical form of Eq. (\ref{star}), and viceversa.  On the same token, the choice of the analytical form of Eq. (\ref{distributiondensity}) for $\psi(t)$ makes the auto-correlation function $\Phi_{\xi}(t)$ lose the analytical form of Eq. (\ref{autocorrelationfunction}). However, using the property of Eq.  (\ref{connection}), it is straigthforward to prove that $\psi (t)$ with the form of  (\ref{distributiondensity}) yields, for the auto-ocorrelation function 
$\Phi _{\xi }(t)$, the following time asymptotic behavior: 
\begin{equation}
\Phi _{\xi }(t)_{t \to \infty}\sim \frac{1}{t^{\beta }},  \label{equilibrium}
\end{equation}
Thus, whatever choice is made, either the analytical form of Eq.  (\ref{distributiondensity}) or the analytical form of Eq.  (\ref{star}),  in both cases the two waiting time distributions maintain the same time asymptotic behavior, with the same 
$\mu$. So do the two different expressions for the equilibrium auto-autocorrelation functions, the time asymptotic equivalence  being the property that matters to study the emergence of fractional derivative.

Herein, using the inverse Laplace transform of  (\ref{mauroisageniustoo}) we
determine the unknown memory kernel $\Phi (t)$, making it possible to
discuss how to express the GME in terms of fractional derivatives. The case
where $2<\mu <3$ is compared to the recent work of Sokolov and Metzler  \cite{sokolov}. We find that the index of the fractional derivative is $3-\mu $,
rather than $\mu -2$, as predicted by Sokolov and Metzler. We prove that
this difference in index is due to the fact that we adopt a stationary
condition, while Sokolov and Metzler do not. We also prove that in the case
of a finite, rather than infinite age, our GME makes a transition from the $%
\left( 3-\mu \right) $-th to the $\left( \mu -2\right) $-th order. The
stationary case becomes stable only in the limiting case of infinite age.
Thus, on the one hand we shed light on the meaning of the work of Allegrini 
{\em et al.}  \cite{gerardo}, which is proven to be a subordination to a
Markov master equation through the stationary distribution of first exit
times. On the other hand, we extend the approach to systems of any age and
reveal the phenomenon of a continuous time random walk with rejuvenation. To
accomplish this dual role we rely heavily on the results recently obtained
by Barkai  \cite{barkai} and, to a lesser extent, the results of Allegrini 
{\em et al.}  \cite{gerardo}. However, this allows us to reveal some aspects
of aging and rejuvenation dependent on the order of fractional derivatives,
which were not previously identified.

\section{The inverse Laplace transform of the memory kernel}

To establish the form of the unknown memory kernel $\Phi (t)$, we make a few
preliminary observations. First of all, we note that through Eq. (\ref
{mauroisageniustoo}) we establish a direct connection with the correlation
function $\Phi _{\xi }(t)$ and that this correlation function is, in turn,
directly related to the waiting time distribution $\psi ^{*}(t)$, through
Eq. (\ref{geiselcrucial1}). Thus, with no loss of generality for the reasons illustrated in Section 1, it is
convenient to refer ourselves to $\psi ^{*}(t)$ rather than to $\psi (t)$.
For simplicity, we set $T^{*}=1$ throughout the section. Thus, the Laplace
transform of the autocorrelation function is  \cite{bologna}:

\begin{equation}
\hat{\Phi}_{\xi }\left( u\right) =\frac{\Gamma \left( 1-\beta \right) }{%
u^{1-\beta }}\left( e^{u}-E_{\beta -1}^{u}\right) ,\text{ }  \label{lapauto}
\end{equation}
where $0<\beta <1$, given the fact that we are considering $2<\mu <3$, and $%
E_{\beta -1}^{u}$ is the generalized exponential function  \cite{brucemauro}$%
. $ Thus, $\hat{\Phi }_{\xi }\left( u\right) $ diverges as $%
u\rightarrow 0$ and Eq. (\ref{lapcorr}) yields $\hat{\Phi }%
\left( 0\right) =0$. We explore the opposite limit, $u\rightarrow \infty $
using Eq.  (\ref{gerardoisagenius}), which yields $\hat{\Phi}\left( u\right)
=\frac{1}{\tau }=\frac{1}{2\tau ^{*}}$. In the time representation, the
latter limit is equivalent to $\Phi \left( t\right) \approx \frac{\delta
\left( t\right) }{2\tau ^{*}}$ for $t\to 0$. Therefore, we segment the
Laplace transform of the GME memory kernel into two parts as follows 
\begin{equation}
\hat{\Phi}\left( u\right) =\frac{1}{2\tau ^{*}}+\hat{\Phi}_{a}\left(
u\right) .  \label{decom1}
\end{equation}
The first term models the short-time limit, while the second term is
responsible for the long-time behavior. In the time representation we have 
\begin{equation}
\Phi \left( t\right) =\frac{\delta \left( t\right) }{2\tau ^{*}}+\Phi
_{a}\left( t\right) .  \label{decomposition}
\end{equation}
Note that this division of the memory kernel into a white-noise contribution
and a slow term corresponds to a similar partition made by Fuli\'{n}ski  \cite{fulinski}.

Thus, for the time evolution equation of the correlation function of $\xi
(t) $, we derive the following equation 
\begin{equation}
\frac{d\Phi _{\xi }\left( t\right) }{dt}=-\frac{1}{\tau ^{*}}\Phi _{\xi
}\left( t\right) -2\int\limits_{0}^{t}\Phi _{a}\left( t-t^{\prime }\right)
\Phi _{\xi }\left( t^{\prime }\right) dt^{\prime }.  \label{58a}
\end{equation}
Using Eq.  (\ref{correlationfunction}) and substituting into it the explicit
expression of $\tau ^{*}$ as a function of $\mu $, after some algebra, we
obtain

\begin{eqnarray}\label{58b}
\left( \frac{T}{t+T}\right) ^{\beta }\frac{\beta t}{T\left( t+T\right) }%
&=&\Phi _{\xi }\left( t\right) \frac{\beta t}{T\left( t+T\right) }\\
\nonumber &=&-2\int\limits_{0}^{t}\Phi _{a}\left( t-t^{\prime }\right) \Phi _{\xi
}\left( t^{\prime }\right) dt^{\prime }.  
\end{eqnarray}
The two terms on the left-hand side Eq.  (\ref{58b}) are positive. Due to the
negative sign on the right-hand term of this equation we conclude that it
might well be that $\Phi _{a}\left( t\right) $ is always negative.

Let us concentrate on the case $\beta<1$: using the autocorrelation function 
$\Phi _{\xi }\left( t\right) $ of Eq.  (\ref{autocorrelationfunction}) (with
T* = 1) and using the change of time variable $t^{\prime }+1\to t^{\prime }$%
, we rewrite Eq.  (\ref{58b}) in the form:

\begin{equation}\label{beta0}
\begin{split}
&-2\int\limits_{0}^{t+1}\Phi _{a}\left( t+1-t^{\prime }\right) \frac{1}{t^{\prime }{}^{\beta }}dt^{\prime }+2\int\limits_{0}^{1}\Phi _{a}\left(t+1-t^{\prime }\right) \frac{1}{t^{\prime }{}^{\beta }}dt^{\prime }\\
&=\frac{\beta t}{\left( t+1\right) ^{\beta +1}}.  
\end{split}
\end{equation}

\begin{figure}
\includegraphics[width=4.5cm, height=6.2cm,angle=270]{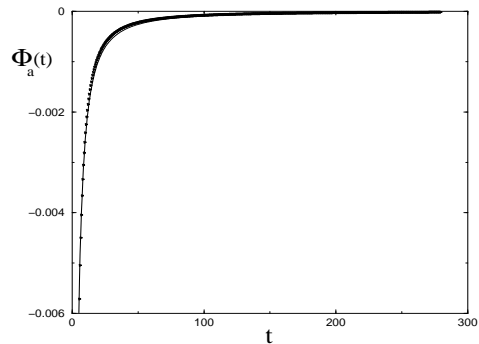}
\caption{\label{fig1}The slow component of the  memory kernel $\Phi(t)$, $\Phi_{a}(t)$, as a function of time. The black dots denote the result of the numerical inversion  of the expression in Laplace transform  resulting from  Eqs. (\ref{gerardoisagenius}) and (\ref{decom1}) for $\beta=0.5$, the continuous line is 
the  analytical approximation given by Eq. (\ref{beta0ris}).}
\end{figure}

In the limiting case $t\rightarrow \infty $ we neglect the second term on
the left-hand side of this equation. This is a natural consequence of the
assumption that the memory kernel must tend to zero with a negative tail, as
an inverse power law. With this assumption it is straightforward to prove
that the modulus of the first term becomes much larger than that of the
second-term on left-hand side of this equation. The consequences of this
crucial assumption are supported by the numerical results depicted
in Fig.  (\ref{fig1}). With this assumption Eq. (\ref{beta0}) simplifies to 
\begin{equation}
-2\int\limits_{0}^{t}\Phi _{a}\left( t-t^{\prime }\right) \frac{1}{t^{\prime
}{}^{\beta }}dt^{\prime }=\frac{\beta t}{\left( t+1\right) ^{\beta +1}},
\label{beta01}
\end{equation}
which can be solved by means of the fractional calculus  \cite{brucemauro}.
We use the Riemann-Liouville (RL) definition of the fractional integral

\begin{equation}
D_{t}^{-q}\left[ f\left( t\right) \right] =\frac{1}{\Gamma \left( q\right) }%
\int_{0}^{t}\frac{f\left( t^{\prime }\right) dt^{\prime }}{\left(
t-t^{\prime }\right) ^{1-q}},  \label{RLI}
\end{equation}
which is the anti-derivative of the fractional derivative with order $q$,
with $q<1.$ Consequently for $\beta <1$ we can express  (\ref{beta01}) in
terms of the RL fractional integral

\[
D_{t}^{\beta -1}\left[ \Phi _{a}\left( t\right) \right] =-\frac{1}{2\Gamma
\left( 1-\beta \right) }\frac{\beta t}{\left( t+1\right) ^{\beta +1}}, 
\]
so that inverting this equation we have the formal expression for the slow
part of the memory kernel

\begin{eqnarray}
\Phi _{a}\left( t\right) &=&-\frac{1}{2\Gamma \left( 1-\beta \right) }%
D_{t}^{1-\beta }\left[ \frac{\beta t}{\left( t+1\right) ^{\beta +1}}\right] 
\nonumber \\
&=&-\frac{1}{2\Gamma \left( 1-\beta \right) }D_{t}^{1-\beta }\left[ \beta
t\Phi _{\xi ,\beta +1}\left( t\right) \right] .  \label{beta0fraz}
\end{eqnarray}
We denote the correlation function with the inverse power $\left( \beta
+1\right) $ by the symbol $\Phi _{\xi ,\beta +1}\left( t\right) $. Carrying
out the required calculations on the right-hand side of  (\ref{beta0fraz}),
we obtain, see for example pg.90 of West {\em et al.}  \cite{brucemauro}, 
\begin{equation}\label{beta0ris}
\Phi _{a}\left( t\right) =-\frac{\beta  }{2\Gamma(1-\beta) \Gamma(1+\beta)}\frac{t^{\beta }}{\left( t+1\right) ^{2}},\text{ } \beta<1 
\end{equation}
A comparison with a numerical inversion of the kernel is shown in Fig.  (\ref{fig1}). 

For the sake of completeness, it is worth noticing that we can proceed in a
similar way also in the case $\beta >1$. In this case we find, for the
contribution $\Phi _{a}(t)$ of the GME, the following time asymptotic
behavior 
\begin{equation}
\Phi _{a}\left( t\right) =-\frac{\beta \left( \beta -1\right) }{2}\frac{t}{%
\left( t+1\right) ^{\beta +1}},\text{ }\beta >1.  \label{beta0ris2}
\end{equation}

\section{The emergence of fractional operators}

In this section we show that in the two-site case we are discussing, the GME
has the form of a transport equation, with two terms on the right-hand side.
The first has the form afforded by the ordinary master equation and
consequently satisfies the Onsager Principle, giving a relaxation dependent
on the average waiting time of CTRW. The second term corresponds to a
fractional derivative in time, and extends the Onsager Principle to the case
of a relaxation with a fat tail. To obtain these results, we use what we
have learned in the preceding section.

First of all, since we are dealing with the two-site case, using the form of 
{\bf M} and {\bf K} matrices, with $z=x=-1$ and $y=1$, we rewrite Eq.  (\ref
{kenkre}) in the following form:

\begin{eqnarray}
&\frac{d}{dt}p_{1}(t)=\int\limits_{0}^{t}\Phi \left( t-\tau \right) \left(
p_{2}(\tau)-p_{1}(\tau)\right) d\tau &,  \label{prob1} \\
&\frac{d}{dt}p_{2}(t)=\int\limits_{0}^{t}\Phi \left( t-\tau \right) \left(
p_{1}(\tau)-p_{2}(\tau)\right) d\tau &.  \label{prob2}
\end{eqnarray}
The  memory kernel $\Phi $ is related to the autocorrelation function of the
dichotomous variable $\Phi _{\xi }$ through Eq.  (\ref{mauroisageniustoo}).
Inserting Eq. (\ref{mauroisageniustoo}) into the Laplace transform of the set of the
two-site dynamical equations, solving the resulting set of equations, and
taking the corresponding inverse Laplace transforms yields the solution 
\begin{eqnarray}
&p_{1}(t)=\frac{1}{2}\left[ 1-\Phi _{\xi }\left( t\right) \left(
p_{2}(0)-p_{1}(0)\right) \right] &  \label{probsol1} \\
&p_{2}(t)=\frac{1}{2}\left[ 1+\Phi _{\xi }\left( t\right) \left(
p_{2}(0)-p_{1}(0)\right) \right] &.  \label{probsol2}
\end{eqnarray}
Note that these solutions can be combined to yield the generalized Onsager
Principle given by  (\ref{ctrw8}) in terms of the difference in the
probabilities.

We now want to find a formal equation of evolution for the probabilities
involving fractional operators. We know from the preceeding section that: 
\begin{equation}
\Phi \left( t\right) =\frac{\delta \left( t\right) }{2\tau ^*}+\Phi
_{a}\left( t\right)  \label{kerr1}
\end{equation}
with $\tau^* =\frac{T*}{\mu -2}=\frac{1}{\beta }$, thanks to the fact that
we set $T*=1$. Substituting the decomposition of the memory kernel into Eq. (%
\ref{prob1}), we obtain 
\begin{equation}
\frac{dp_{1}(t)}{dt}=\frac{p_{2}(t)-p_{1}(t)}{2\tau^* }+\int\limits_{0}^{t}%
\Phi _{a}\left( t-\tau \right) \left( p_{2}(\tau)-p_{1}(\tau)\right) d\tau .
\label{fracder}
\end{equation}
Writing $\Phi _{a}\left( t\right) $ as the derivative of an as yet
unspecified function $f(t)$, and using the property

\begin{equation}
\frac{d}{dt}\int\limits_{0}^{t}f(t-\tau )g(\tau )d\tau
=f(0)g(t)+\int\limits_{0}^{t}f^{\prime }(t-\tau )g(\tau )d\tau
\label{convder}
\end{equation}
with $f^{\prime }(t)=\frac{d}{dt}f(t)$, we obtain

\begin{eqnarray}
\frac{dp_{1}(t)}{dt}&=&\frac{p_{2}(t)-p_{1}(t)}{2\tau ^{*}}-f(0)\left( p_{2}(t)-p_{1}(t)\right) \\
\nonumber &+&\frac{d}{dt} \int\limits_{0}^{t}f(t-\tau )\left( p_{2}(\tau )-p_{1}(\tau )\right) d\tau
 .  \label{fracder2}
\end{eqnarray}
Also we found that in the case $0<\beta <1$ the asymptotic behavior of the
memory kernel is expressed by 
\begin{eqnarray}
\Phi _{a}\left( t\right) &\approx& -\frac{\beta}{2 \Gamma(1-\beta) \Gamma(1+\beta)}\frac{t^{\beta }}{\left( t+1\right) ^{2}}\\
\nonumber&\approx& -\frac{\beta}{2 \Gamma(1-\beta) \Gamma(1+\beta)}t^{\beta -2}
  \qquad t\to
\infty ,\,\,T^{*}=1.  \label{betasym}
\end{eqnarray}
Rewriting Eq.  (\ref{betasym}) as: 
\begin{equation}
\Phi _{a}\left( t\right) \approx -\frac{1}{2\Gamma (1-\beta )\Gamma
(\beta )(\beta-1)}\frac{d}{dt}t^{\beta -1},  \label{betasym2}
\end{equation}
we identify $f(t)$ with $\frac{1 }{2(1-\beta)\Gamma (1-\beta )\Gamma (\beta )}t^{\beta -1}$ for $t\to \infty $. Choosing $f(0)=0$ and using the properties of the gamma function, we assign to the time
asymptotic equation of motion the following form 
\begin{eqnarray}
&&\frac{dp_{1}(t)}{dt}=-\frac{p_{1}(t)-p_{2}(t)}{2\tau ^{*}}\\
\nonumber &+&\frac{1}{2 \Gamma (2-\beta )\Gamma (\beta )}\frac{d}{dt}\int\limits_{0}^{t}(t-\tau
)^{\beta -1}\left( p_{2}(\tau )-p_{1}(\tau )\right) d\tau  \label{betasym3}
\end{eqnarray}
or, in terms of a RL fractional integral  (\ref{RLI}), 
\begin{equation}
\frac{dp_{1}(t)}{dt}=-\frac{p_{1}(t)-p_{2}(t)}{2\tau ^{*}}-\frac{1}{2 \Gamma (2-\beta )}D_{t}^{1-\beta }\left[ p_{1}(t)-p_{2}(t)\right] .
\label{betasym3b}
\end{equation}
The same procedure applied to the equation of motion for $p_{2}(t)$ yields, 
\begin{equation}
\frac{dp_{2}(t)}{dt}=-\frac{p_{2}(t)-p_{1}(t)}{2\tau ^{*}}+\frac{1}{2\Gamma (2-\beta )}D_{t}^{1-\beta }\left[ p_{1}(t)-p_{2}(t)\right] .
\label{betasym32}
\end{equation}

The difference between these two equations yields

\begin{eqnarray}
\frac{d}{dt}\left[ p_{1}(t)-p_{2}\left( t\right) \right] &=&-\frac{%
p_{1}(t)-p_{2}(t)}{\tau ^{*}}\\
\nonumber &-&\frac{1}{\Gamma (2-\beta )}%
D_{t}^{1-\beta }\left[ p_{1}(t)-p_{2}(t)\right]  \label{difference}
\end{eqnarray}
clearly showing the two kinds of contributions to the generalized Onsager
Principle. The first term gives the relaxation of the perturbation away from
equilibrium at the macroscopic rate required by Onsager. The second term
gives the additional slow relaxation in the form of the fractional integral.

\section{Aging order}

We have to remark that the condition of Eq. (\ref{lapcorr}) refers to the
stationary condition explicitly considered by Klafter and Zumofen  \cite{stationary}. In this section we prove that there is a connection between a
system's age and the order of the fractional derivative in the relaxation
process. A sign of the dependence of the fractional derivative order on age
is given by the discrepancy between the result of Section 3 and Ref.  \cite{sokolov}. Let us compare Eq.  (\ref{betasym3b}) to Eq.(16) of Ref.  \cite{sokolov}. We obtain the fractional index $1-\beta $ rather than $\beta $ as
in the work of Sokolov and Metzler. In Appendix 1 we prove that our time
asymptotic approach to fractional derivatives, in the non-stationary case
studied by Sokolov and Metzler yields the same index as they obtain  \cite{sokolov}. Thus, the discrepancy between our prediction and the prediction
of Sokolov and Metzler depends on the fact that we consider a condition
consistent with the Onsager principle, whereas Sokolov and Metzler do not.
Furthermore, if the system is not infinitely aged, a sort of rejuvination
process is expected to take place that will lead to the fractional order of
Sokolov and Metzler.

To support our remarks concerning the relation between aging and the order
of fractional operator, here we discuss how to define a waiting time
distribution of any age. The authors of Ref.  \cite{gerardo} have shown that
the waiting time distribution $\psi (t)$ of Eq.  (\ref{distributiondensity})
is obtained from the following dynamic model. A particle moves in an the
interval $I\equiv [0,1]$ driven by the equation of motion 
\begin{equation}
dy/dt=\alpha y^{z},  \label{ourbath}
\end{equation}
with $z>1$. When the particle reaches the border $y=1$, it is injected back
to an initial condition between $y=0$ and $y=1$ with uniform probability.
The age of the CTRW is determined by the distribution of first exit times.
The ordinary CTRW rests on identifying this distribution with $\psi (t)$.
This means that the CTRW is equivalent to assuming that the system is
prepared in a flat distribution at $t=0$, which coincides with the beginning
of the observation process.

Let us discuss now the consequence of beginning the observation a
significant time after the preparation. Let us imagine that the system is
prepared in a flat distribution at a time $t=-t_{a}<0$, and that the
observation begins at $t=0$. This means that the flat distribution begins
producing a sequence of time intervals of size $\tau $, according to the
distribution of Eq.  (\ref{distributiondensity}); more precisely, the time
interval $\tau _{1}$ beginning at $t=-t_{a}$ and ending at $t=-t_{a}+\tau
_{1}$, the time interval $\tau _{2}$ beginning at $t=-t_{a}+\tau _{1}$ and
ending at $t=-t_{a}+\tau _{1}+\tau _{2}$, and so on. The waiting time
distribution of age $t_{a}$, denoted by $\psi _{t_{a}}(t)$, is determined by
the first of these time intervals overlapping with $t>0$. The time length of
that overlap is the time length whose distribution determines $\psi
_{t_{a}}(t)$. We make the assumption that the beginning of the first time
interval overlapping with $t>0$ occurs with equal probability at any point
between $t=-t_{a}$ and $t=0$. The validity of this assumption is discussed
in Appendix 2, which establishes that this assumption is very good for $%
t_{a}\rightarrow 0$ and $t_{a}\rightarrow \infty $. In between the
asymptotic limits the resulting prediction is not exact. However, since it yields simple analytical formulas, we adopt this 
simplifying assumption for any age.
Thus, we have that

\begin{equation}
\psi _{t_{a}}(t)=\frac{\int_{0}^{t}\psi (t+y)dy}{g(t_{a})},
\label{genialideaofgerardo}
\end{equation}
where $g(t_{a})$ is the normalization factor defined by 
\begin{equation}
g(t_{a})\equiv \int_{0}^{t_{a}}\Psi (t^{\prime })dt^{\prime },
\label{normalization}
\end{equation}
and $\Psi (t)$ is the probability that no event occurs throughout the time
interval of length $t$. Using for $\psi (t)$ the explicit form of Eq. (\ref
{distributiondensity}), it is easy to prove that Eq.  (\ref
{genialideaofgerardo}) can be written under the form 
\begin{equation}\label{explicitform}
\psi _{t_{a}}(t)=(\mu -2)\frac{(t+T)^{(1-\mu )}-(t+T+t_{a})^{(1-\mu )}}{%
T^{(2-\mu )}-(t_{a}+T)^{(2-\mu )}}.  
\end{equation}
This formula proves that for $t<<t_{a}$ the index of the distribution is $%
\mu -1$, whereas for $t>>t_{a}$ the index becomes $\mu $. This result for
the age-dependent waiting time distribution function agrees with the
predictions by Barkai  \cite{barkai} and by the authors of Ref. \cite{gerardo}%
. Notice that the formula Eq. (\ref{explicitform}) is equivalent to drawing
the initial condition for $y$ from an aged distribution of this variable.

Here, we are in a position to evaluate the waiting time index at a generic
time, where, we write $\psi _{t{a}}(t)$, as 
\begin{equation}
\psi _{t{a}}(t)=\frac{A(T,t_{a})}{(t+T)^{\mu _{eff}(t)}}.
\label{inversepower}
\end{equation}
Using Eq. (\ref{explicitform}) we arrive at the following formula for the
time dependence of the effective power-law index 
\begin{equation}
\mu _{eff}(t)=-\frac{ln[(t+T)^{(1-\mu )}-(t+T+t_{a})^{(1-\mu )]}]}{ln[t+T]}.
\label{atalltimes}
\end{equation}
\begin{figure}
\includegraphics[width=4.3cm,height=7cm,angle=270]{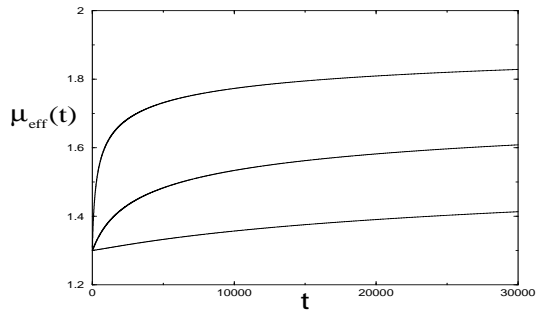}

\caption{\label{fig2}The effective power index $\mu_{eff}(t)$ as a function of time, for $\mu=2.3$. The curves refer, from top to bottom, to $t_{a} = 100, 1000, 10000$.}
\end{figure}
Fig.  (\ref{fig2}) illustrates the regression of the effective power-law
index to $\mu $ with two different ages, and shows clearly that the
regression is slower for older systems. This formula does more than explain
the discrepancy between Eq.  (\ref{betasym3b}) and Eq. (16) of Ref.  \cite{sokolov}. In fact, it shows that it is possible to build a GME that at
short times follows the prescription of our GME and at long times moves into
the basin of attraction of Sokolov and Metzler. This is certainly the case,
if $t_{a}>-\infty $.

This aspect is important and needs a more exhaustive illustration. We note
that the approach of Ref.  \cite{gerardo} can be easily extended to the case
where the distribution of first exit times has a finite age. It is enough to
follow the procedure of Ref.  \cite{gerardo} and to replace the first exit
time distribution with $\psi _{t_{a}}(t)$ rather than with $\psi _{\infty
}(t)\equiv \psi _{t_{a}=-\infty }(t)$, as done in Ref.  \cite{gerardo}. The
result of this procedure yields for the GME the following form for the
Laplace transform of the memory kernel 
\begin{equation}
\hat{\Phi}_{t_{a}}(u)=\frac{u\hat{\psi}_{t_{a}}(u)}{1+\hat{\psi}(u)-2\hat{%
\psi}_{t_{a}}(u)}.  \label{kernelgen}
\end{equation}
It straightforward to prove that for $t_{a}=0$  (\ref{kernelgen}) reduces to 
\begin{equation}
\hat{\Phi}_{t_{a}=0}(u)=\frac{u\hat{\psi}(u)}{1-\hat{\psi}(u)}.
\label{montroll}
\end{equation}
In fact, the general prescription of Eq. (\ref{genialideaofgerardo})
immediately yields $\psi _{t_{a}}(t)=\psi (t)$ for $t_{a}=0$. To derive the
memory kernel corresponding to the infinitely aged condition of Eq.  (\ref{gerardoisagenius}) we have to notice first that Eq. (\ref{genialideaofgerardo}) yields, in accordance with Refs.  \cite{barkai} and 
 \cite{gerardo}, 
\begin{equation}
\psi _{\infty }(t)\equiv \psi _{t_{a}=\infty }(t)=\frac{1}{\tau }%
\int_{t}^{\infty }dt^{\prime }\psi (t^{\prime }).  \label{infinity}
\end{equation}

To illustrate the change of the memory kernel with time, notice that the
Laplace transform of Eq.  (\ref{explicitform}) is: 
\begin{equation}
\hat{\psi}_{t_{a}}(u)=\frac{\hat{\Psi}(u)(1-e^{ut_{a}})+e^{ut_{a}}%
\int_{0}^{t_{a}}e^{-uy}\Psi (y)dy}{g(t_{a})}
\end{equation}
that is: 
\begin{equation}
\hat{\psi}_{t_{a}}(u)=\frac{(1-\hat{\psi}(u))(1-e^{-ut_{a}})+ue^{ut_{a}}%
\int_{0}^{t_{a}}e^{-uy}\Psi (y)dy}{ug(t_{a})} .  \label{laplacewait}
\end{equation}
\begin{figure}
\includegraphics[width=7.8cm,height=4.3cm,angle=0]{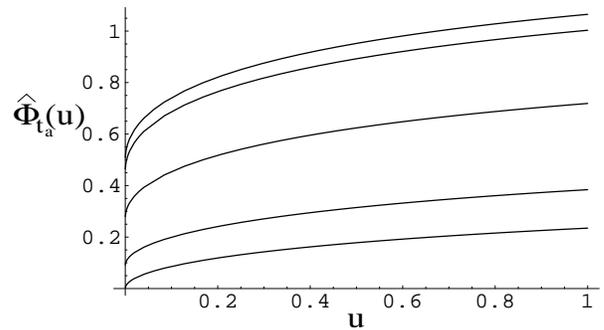}

\caption{\label{fig3}The Laplace transform of the  $t_{a}$-old memory kernel 
$\Phi_{t_{a}}(t)$, $\hat \Phi_{t_{a}}(u)$ as a function of u, for $\mu=2.3$. The curves 
refer, from top to bottom, to  $t_a=0,\ 0.1,\ 1,\ 10,\ \infty$}
\end{figure}

 Let us substitute  (\ref{laplacewait}) into Eq.  (\ref{kernelgen}),
 from which we obtain the Laplace transform of the memory kernel for an
arbitrary age. In Fig.  (\ref{fig3}) we show the Laplace transform of the
memory kernel corresponding to a number of different ages. Moving from the
top to the bottom curve the age increases from the brand new kernel ($%
t_{a}=0 $) to the infinitely aged, or stationary kernel ($t_{a}=\infty $).

\section{Concluding Remarks}

In this paper we establish that an infinitely old system, with the power-law
index in the interval $2\leq \mu \leq 3$, yields the fractional order $%
1-\beta =3-\mu $. The prediction of Ref.  \cite{sokolov}, on the other hand,
yields the fractional order $\mu -2$, corresponding to the brand new
condition. If the system is not infinitely aged, namely $t_{a}<\infty $, the
short-time behavior of the system, for $t<<t_{a}$ is expected to be that of
an aged system. At large observation times, $t>>t_{a}$, rejuvenation begins.
This can be explained using the dynamical model of Eq.  (\ref{ourbath}). In
fact, aging has to do with the slow regression to equilibrium, if it exists,
of the variable $y$, which mimics a real bath slowly regressing to
equilibrium. The aging effects discussed by Barkai  \cite{barkai} corresponds
to $z>2$ and thus to $\mu <2$. In this case the dynamic model under
discussion does not have an invariant distribution, and, consequently, any
observation is done while the bath is drifting towards a condition that will
be never reached. This aging effect affects the form of the first exit time
distribution, whose index is $\mu -1$ rather than $\mu $. However, after the
first exit the trajectories are injected back with a uniform probability,
and thus, all the ensuing jumps are determined by the ordinary waiting time
distribution $\psi (t)$.

It would be desirable to have an equation of motion with a fractional
operator order that changes as a function of time from $3-\mu $ to $\mu -2$.

However, there are technical and conceptual difficulties that make it
difficult, if not impossible, to realize this goal. In fact, according to
the perspective adopted in this paper, the order of the fractional operator
is established using time asymptotic arguments. Thus, if $t_{a} < \infty$ we can associate a time $t << t_{a}$ with the 
order $ord = \mu - 2$, of the fractional derivative. This is so because 
for $t<<t_{a}$ the calculations would be virtually equivalent to 
those done in Appendix I. It is not clear how to proceed when $t$ is is 
of the order of $t_{a}$, this being the first reason why assigning a 
fractional derivative order to any time might be difficult.
There also exists a physical reason that might make it impossible to move
from the order $3-\mu $ to $\mu -2$. Physically this extremely extended
transition process might involve a mixture of fractional derivatives of
different orders. It has been shown  \cite{gianmarco} that the L\'{e}vy walk
does not have well-defined scaling, due to aging effects. Similarly, the
adoption of a fractional time derivative, with time-dependent order, might
be inadequate to explore the regime of transition from the order $3-\mu $ to
the order $\mu -2$. In conclusion, the fractional operator, and its order,
reflects a stable condition, of a brand new or infinitely old system. The
regime of transition from the dynamic to any of these two thermodynamic
regimes, and the regime of transition from the earlier to the latter
thermodynamic regime is not yet a fully understood physical condition, an
issue calling for further investigation.

It is interesting to notice that, even if we select $\mu >2$, and
consequently we adopt a condition compatible with the stationary condition,
the effective index of the first exit distribution is located in the
non-stationary region, if $\beta <1$. This is probably the reason why the
memory kernel seems to share the same properties as those adopted by Ref.%
 \cite{researchpaper3,lutz,pottier} to produce sub-diffusion. Notice that the
baths used by Lutz  \cite{lutz} and Pottier  \cite{pottier} have properties
quite different from the subordination perspective of Ref.  \cite{researchpaper3}, even the relaxation process stemming from subordination 
 \cite{researchpaper3} is quite similar to that produced by the non-ohmic
baths of Lutz and Pottier. We hope that the present work might help
understanding the connection between the two perspectives. This is another
subject for future research.

\emph{Acknowledgments}  G.A and P. G. gratefully acknowledge financial support from ARO through Grant DAAD190210037

${}\newline
$ \setcounter{equation}{0} \renewcommand{\theequation}{A-\arabic{equation}} %
\appendix{\bf APPENDIX I} \newline
\vspace{0.8cm}

In this Appendix we show how to obtain the order of the fractional operator
in the GME for $1<\mu <3$, using the Laplace transform form for $\Phi (t)$
given by Eq. ( \ref{brandnew}). We are referring ourselves to $\psi(t)$ rather than $\psi^{*}(t)$. Thus, for the reasons pointed out in Section 2, we adopt the analytical expression of Eq.  (\ref{distributiondensity}). 

We note that for $u\to \infty $ we get a
finite value: $\hat{\Phi}\left( \infty \right) =\frac{\mu -1}{T}$
corresponding to $\psi \left( 0\right) $ in $t$-space. As done in Section 2,
we separate the kernel into two contributions: $\Phi (t)=\frac{\mu -1}{T}%
\delta (t)+\Phi _{a}(t)$, and insert the Laplace transform of separation
into Eq.  (\ref{brandnew}), to write:

\begin{eqnarray}\label{prenonstaz}
\hat{\Phi}_{a}\left( u\right) &=&\frac{u\hat{\psi} (u) -(\mu -1)+(\mu
-1)\hat{\psi} ( u) }{1-\hat{\psi} ( u) }\\
\nonumber
 &=&\frac{\hat{\psi} _{D}\left(
u\right) +(\mu -1)\hat{\psi} \left( u\right) }{1-\hat{\psi} \left( u\right) }.
\end{eqnarray}
We introduce $\hat{\psi} _{D}\left( u\right) $ is the Laplace transform of the
distribution's derivative $\psi ^{\prime }\left( t\right) =-\mu (\mu -1)%
\frac{T^{\mu -1}}{(T+t)^{\mu +1}}$; using the Laplace transform of an
inverse power law and substituting it into Eq.  (\ref{prenonstaz}) we have:

\begin{equation}\label{nonstaz2}
\begin{split}
\hat{\Phi}_{a}\left( u\right)&=\frac{(\mu -1)^2 \Gamma (1-\mu) ( e^{u}-E_{\mu -1}^{u})}{u^{1-\mu
}-\left( \mu -1\right) \Gamma \left( 1-\mu \right) \left( e^{u}-E_{\mu
-1}^{u}\right)}\\
&-\frac{\mu(\mu-1) \Gamma \left( -\mu \right) u\left( e^{u}-E_{\mu }^{u}\right) }{u^{1-\mu
}-\left( \mu -1\right) \Gamma \left( 1-\mu \right) \left( e^{u}-E_{\mu
-1}^{u}\right)},  
\end{split}
\end{equation}
where, as usual, for simplicity we have set $T=1$. Anticipating the
convolution form of the solution we cross-multiply to obtain:
\begin{equation}\label{nonstaz3}
\begin{split}
\left[ u^{1-\mu } - \left( \mu \right. \right. &-\left.\left. 1\right) \Gamma \left( 1-\mu \right) \left(
e^{u}-E_{\mu -1}^{u}\right) \right] \hat{\Phi}_{a}\left( u\right) \\
 &=(\mu-1)\left( (\mu-1)\Gamma \left( 1-\mu \right) \left( e^{u}-E_{\mu-1}^{u}\right) \right.\\
 \left. \right. &-\left.\mu \Gamma ( -\mu) u \left( e^{u}-E_{\mu }^{u}\right)\right .  
\end{split}
\end{equation}
Using the following relation \cite{brucemauro} 
\begin{equation}\label{extralap}
\int\limits_{0}^{\infty }\frac{t^{\alpha }}{t+a}\exp \left[ -ut\right] dt=%
\frac{\pi a^{\alpha }}{\sin \pi \alpha }\left( E_{\alpha
}^{ua}-e^{ua}\right) ,\,\alpha >-1  
\end{equation}
and setting $\alpha =\mu -1$ we construct

\begin{eqnarray}\label{nonstaz4}
R(t)&=&\frac{1}{\Gamma \left( \mu -1\right) }\int\limits_{0}^{t}t^{\prime }{}^{\mu
-2}\Phi _{a}(t-t^{\prime })dt^{\prime }\\
\nonumber &-&\frac{\sin \pi \mu }{\pi }\left( \mu
-1\right) \Gamma \left( 1-\mu \right) \int\limits_{0}^{t}\frac{t^{\prime
}{}^{\mu -1}}{t^{\prime }+1}\Phi _{a}(t-t^{\prime })dt^{\prime },
\end{eqnarray}
where $R(t)$ is the inverse Laplace transform of right side of Eq.  (\ref
{nonstaz3}). Using the well known recurrence relation of Gamma function, we
can combine terms in Eq.  (\ref{nonstaz4}) to obtain

\begin{equation}
\label{nonstaz4bis}
\begin{split}
&R(t)=\frac{1}{\Gamma \left( \mu -1\right) }\int\limits_{0}^{t}t^{\prime }{}^{\mu
-2}\left( 1-\frac{t^{\prime }}{t^{\prime }+1}\right) \Phi _{a}(t-t^{\prime
})dt^{\prime }\\
&=\frac{1}{\Gamma \left( \mu -1\right) }\int\limits_{0}^{t}%
\frac{t^{\prime }{}^{\mu -2}}{t^{\prime }+1}\Phi _{a}(t-t^{\prime
})dt^{\prime }.  
\end{split}
\end{equation}
Let us consider first the case $\mu >2$, where, in the limit of $t\gg 1$
(equivalent to $t\gg T$), we can write Eq.  (\ref{nonstaz4bis}) to a good
approximation  
\begin{equation}\label{mu<3}
R(t)=\frac{1}{\Gamma \left( \mu -1\right) }\int\limits_{0}^{t}t^{\prime }{}^{\mu-3}\Phi _{a}(t-t^{\prime })dt^{\prime }=R(t).  
\end{equation}
Going back to the Laplace transform representation, we obtain the simplier
expression

\begin{equation}\label{sol1}
\begin{split}
\frac{\Gamma \left( \mu -2\right) }{\Gamma \left( \mu -1\right) }\frac{\hat{\Phi}_{a}\left( u\right) }{u^{\mu -2}}&=\left( \mu -1\right) ^{2}\Gamma
\left( 1-\mu \right) \left( e^{u}-E_{\mu -1}^{u}\right)\\
 &-\mu \left( \mu
-1\right) \Gamma \left( -\mu \right)u\left( e^{u}-E_{\mu }^{u}\right) ,
\end{split}
\end{equation}
which after a little algebra yields for $\hat{\Phi}_{a}\left( u\right) $:

\begin{equation}\label{sol2}
\begin{split}
\hat{\Phi}_{a}\left( u\right) =&\left( \mu -1\right) \left( \mu -2\right)
 \cdot \left[\rule{0 cm}{0.65 cm} \left( \mu -1\right)\Gamma(1-\mu)\frac{1}{u}\frac{e^{u}-E_{\mu -1}^{u}}{u^{1-\mu }} \right.\\
& \left. -\mu \Gamma(-\mu) \frac{1}{u}\frac{\left( e^{u}-E_{\mu}^{u}\right) }{u^{-\mu }}\rule{0 cm}{0.65 cm}\right] .  
\end{split}
\end{equation}
So that using the inverse Laplace transforms we obtain the corresponding
expression in the time representation:

\begin{equation}  \label{sol3}
\Phi_a \left(t \right)= \left(\mu-1 \right)\left(\mu-2 \right) \left[\frac{1%
}{(1+t)^{\mu }} -\frac{1}{(1+t)^{\mu -1}} \right].
\end{equation}
In the time asymptotic limit we get

\begin{equation}\label{sol3bis}
\Phi _{a}\left( t\right) \propto \frac{1}{t^{\mu -1}},  
\end{equation}
corresponding to fractional operator of index $\beta =\mu -2$. For the sake
of completeness, we also give the expression for the GME kernel in the case $%
\mu <2$. Proceeding as done earlier  we obtain:

\begin{equation}\label{apprt}
\Phi \left( t\right) \approx -\frac{\sin \pi \mu }{\pi }\frac{t^{\mu -2}}{%
\left( t+1\right) ^{2}}\left[ \mu -1+\left( \mu -2\right) t\right].
\end{equation}
Finally, we want to point out that the expressions we are proposing 
refer to values of mu which are not integer. We are exploring the 
interval $[1,3]$. Thus the the expressions we are proposing become 
questionable for $\mu = 2$. To get the proper expression for $\mu = 2$ we 
have to apply to study  expressions like those of Eq. (\ref{prenonstaz}) and (\ref{sol1}) 
at $\mu = 2 + \epsilon$, do a Talylor series expansion around $\mu =2$, and 
assign to $\mu = 2$ the limiting values reached for $\epsilon \to
0$.


${}\newline
$
 \setcounter{equation}{0} \renewcommand{\theequation}{B-\arabic{equation}} %
\appendix{\bf APPENDIX II} \newline
\vspace{0.8cm}

This Appendix is devoted to establishing the accuracy of Eq. (\ref
{genialideaofgerardo}), and consequently the validity of the assumption that
the beginning of the first laminar region overlapping with $t>0$ is
uniformely distributed between $t=-t_{a}$ and $t=0$. Let us set $t_{a}=0$.

The exact expression for $\psi _{t_{a}}(t)$ is 
\begin{equation}
\psi _{t_{a}}(t)=\int_{0}^{t_{a}}dxG(t_{a}-x)\psi (t+x),  \label{exact}
\end{equation}

where 
\begin{eqnarray}\label{g}
&&G(t)\equiv \delta(t)+\psi(t)+\sum_{n=2}^{\infty }\int_{0}^{t}d\tau _{1}\psi (\tau_{1})\\
\nonumber && \cdot \int_{\tau _{1}}^{t}d\tau _{2}\psi (\tau _{2}-\tau _{1})...\int_{\tau_{n-2}}^{t}d\tau _{n-1}\psi (t-\tau _{n-1}).  
\end{eqnarray}
\begin{figure}[h!]
\includegraphics[width=5.2cm,height=7.6cm,angle=270]{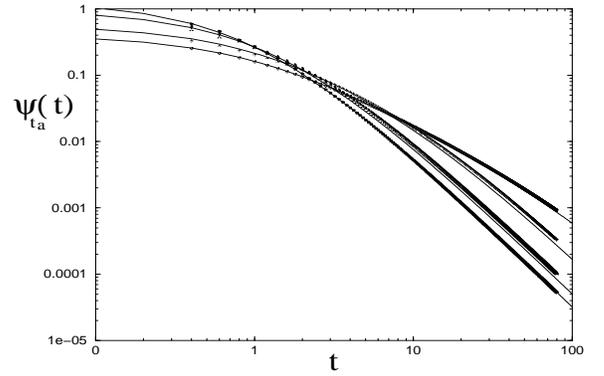}
\caption{\label{fig4}
The waiting time distribution $\psi_{t{a}}(t)$ as a function ot time $t$. 
The dots denote the exact values, and the lines the prediction of Eq. (\ref{explicitform}) for $\mu=2.3, T=1$. Moving from bottom to top in the right-hand portion of the $t_a=0.01, 1, 10, 100$}.
\end{figure}

It is straightforward to find the Laplace transform of $G(t)$. This is given
by 
\begin{equation}
\hat{G}(u)=\sum_{n=0}^{\infty }\hat{\psi}(u)^n=\frac{1}{1-\hat{\psi}(u)}.
\label{laplacetransformofg}
\end{equation}
Thus, the Laplace transform of  (\ref{exact}) respect to $t_{a}$ reads: 
\begin{equation}
\psi _{s_{a}}(t)=\frac{1}{1-\psi (s_{a})}e^{s_{a}t}\left[ \psi
(s_{a})-\int_{0}^{t}e^{-s_{a}y}\psi (y)dy\right]   \label{exactlaplace}
\end{equation}
By the numerical anti-Laplace transforming Eq. (\ref{exactlaplace}) we
evaluate the time dependence of the exact waiting time distribution of age $%
t_{a}$, Eq. (\ref{exact}). In Fig.  (\ref{fig4}) we compare the exact
prediction, evaluated numerically, to the heuristic expression of Eq.  (\ref{genialideaofgerardo}). We find that at short and large values of $t_{a}$
these two expressions coincide. In the intermediate region they do not.
Nevertheless, we think the agreement between the two expressions is
satisfactory enough for the purpose of this paper.

\end{document}